\documentclass[12pt,a4paper,oneside]{article}
\usepackage[latin1]{inputenc}
\usepackage{amsmath}
\usepackage{amsfonts}
\usepackage{amssymb}
\usepackage{graphicx}
\usepackage{caption}
\usepackage{subcaption}
\usepackage{color}
\usepackage{chngcntr}
\usepackage{authblk}
\counterwithout{figure}{section}
\author{Erik D\'iaz-Bautista and David J. Fernandez C.}
	\affil{\small Physics Department, Cinvestav, P.O. Box 14-740, 07000 Mexico City, Mexico}
	\date{\today}
\title{Graphene coherent states}
\begin{document}
		\maketitle
		
		\begin{abstract}
			In this paper we will construct the coherent states for a Dirac electron in graphene placed in a constant homogeneous magnetic field which is orthogonal to the graphene surface. First of all, we will identify the appropriate annihilation and creation operators. Then, we will derive the coherent states as eigenstates of the annihilation operator, with complex eigenvalues. Several physical quantities, as the Heisenberg uncertainty product, probability density and mean energy value, will be as well explored.
		\end{abstract}
	
	\section{Introduction}
	Graphene is a single layer of carbon atoms arranged in a hexagonal honeycomb lattice, which is the basic structural element of other allotropes including graphite, charcoal, carbon nanotubes and fullerenes.
	
\begin{figure}[h]
\centering
\begin{minipage}[h]{0.45\textwidth}
	\includegraphics[width=\textwidth]{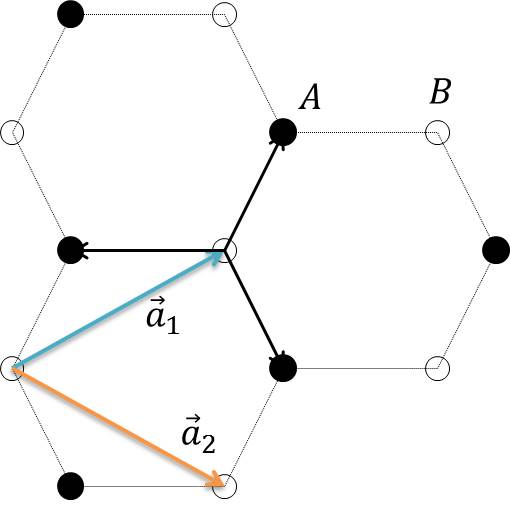}
\end{minipage}
%\hfill
\hspace{1cm}
\begin{minipage}[h]{0.45\textwidth}
	\includegraphics[width=\textwidth]{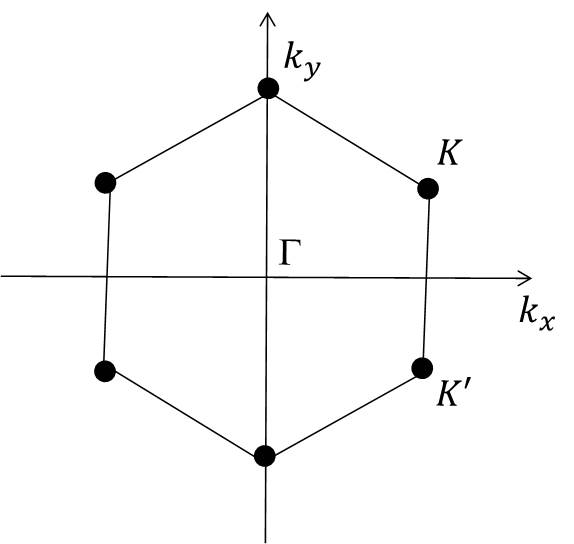}
\end{minipage}
\caption{Left: Lattice structure of the graphene, where the sublattices are labeled by A and B. Right: Brillouin zone for the graphene. The Dirac cones appear at the $K$ and $K'$ points.}
\label{fig:lattice}
\end{figure}

	Graphene is a zero-gap semiconductor, because its conduction and valence bands meet at the Dirac points which are six locations in momentum space, on the edge of the Brillouin zone, divided into two non-equivalent sets of three points, typically labeled as $K$ and $K'$. By contrast, for traditional semiconductors the point of primary interest is denoted as $\Gamma$, where the momentum is zero \cite{cooper,novoselov04,novoselov05,zhang05,nair08,enoki} (see Figure \ref{fig:lattice}).

	Thus, even neglecting their spin, at low energies the electrons can be described by an equation that is formally equivalent to the massless Dirac equation:
	\begin{equation}\label{1}
		-i\hbar\, v_F\, \vec \sigma \cdot \nabla \Psi(\mathbf{r})\,=\,E\Psi(\mathbf{r}).
	\end{equation}
	Here $v_F \sim 10^6$ m/s (.003 $c$) is the Fermi velocity in graphene, which replaces the velocity of light in Dirac theory, $\vec{\sigma}$ is the vector of Pauli matrices, $\Psi(\mathbf{r})$ is the two-component wave function of the electrons and $E$ is its energy \cite{novoselov}.

As a consequence, the electrons and holes are called Dirac fermions; their appearance was predicted in the silicene, germanene or dichalcogenides \cite{semenoff,cahangirov09,fleurence12,vogt12,xiao12,andreev64,volovik92} and emerge naturally from the tight-binding model for a generic hexagonal lattice in the low-energy regime \cite{hasegawa06}, as shown in Figure \ref{fig:lattice}. This pseudo-relativistic description is restricted to the chiral limit, i.e., to vanishing rest mass leading to interesting additional features: when Dirac fermions are compared with ordinary electrons placed in magnetic fields, their behavior leads to new physical phenomena such as the anomalous integer quantum Hall effect, the Zitterbewegung and the Klein paradox \cite{semenoff,stander}.
	
It is important to stress that graphene belongs to a class of systems in condensed matter for which the low-energy quasi-particles behave like massless or massive Dirac fermions. These systems are known as Dirac materials in the literature \cite{wehling14}.

As we shall see below, under particular physical conditions, a problem similar to that considered in \cite{ed16} arises naturally. Due to this, it seems obvious the need to build up the coherent states for the graphene, and then to analyze their properties.
	
In order to do that, this paper is organized as follows. In section \ref{sec:level2} the Dirac-Weyl equation will be introduced and the physical problem to be considered will be briefly discussed. In section \ref{sec:level3} the annihilation operator associated to the system will be defined, and the corresponding coherent states will be constructed as eigenstates of that operator. We will analyze as well several physical quantities for these states. Our conclusions will be presented in section \ref{sec:level4}.
	
	\section{\label{sec:level2}Dirac-Weyl equation}
	
Let us suppose now that the graphene is placed in a static magnetic field which is orthogonal to the material surface (the $x-y$ plane) \cite{novoselov,kuru,midya14}. The interaction of a Dirac electron with such a field close to a Dirac point $K$ in the Brillouin zone is described by the Dirac-Weyl equation, which is obtained by replacing in Eq.~(\ref{1}) the momentum operator $\vec{p}=-i\hbar\,\nabla$ by $\vec{p}+e\vec{A}/c$, leading to:
	\begin{equation}\label{2}
		v_F\, \vec{\sigma}\cdot\left(\vec{p}+\frac{e\vec{A}}{c}\right)\Psi(x,y)\,=\,E\Psi(x,y),
	\end{equation}
	where $-e$ is the charge of the electron. Landau gauge is conveniently chosen, with the vector potential given by $\vec{A}=A(x)\hat{e}_y$ and $\vec{B}=\nabla\times\vec{A}=B(x)\hat{e}_z$, $B(x) = A'(x)$. Taking into account the translational invariance along the $y$ direction, the two-component spinor $\Psi(x,y)$ is expressed as:
	\begin{equation}\label{3}
		\Psi(x,y)=e^{iky}\left(
		\begin{array}{c}
			\psi^+(x) \\
			i\psi^-(x) \\
		\end{array}
		\right),
	\end{equation}
being $k$ the wave number in the $y$ direction and $\psi^{\pm}(x)$ describing the electron amplitude on two adjacent sites in the unit cell of graphene. Thus, the Dirac-Weyl equation (\ref{2}) yields two coupled first-order linear differential equations
	\begin{equation}\label{4}
		\left(\pm\frac{d}{dx}+\frac{e}{c\hbar}A(x)+k\right)\psi^\mp(x)=\frac{E}{\hbar\,v_F}\psi^\pm(x),
	\end{equation}
which can be easily decoupled into two Schrödinger equations $H^\pm\psi^\pm(x)=\mathcal{E}\psi^\pm(x)$, where
	\begin{equation}\label{5}
		H^\pm=-\frac{d^2}{dx^2}+V^\pm, \quad V^\pm=\left(\frac{eA(x)}{c\hbar}+k\right)^2\pm\frac{e}{c\hbar}\frac{dA(x)}{dx}, \quad \mathcal{E}=\frac{E^2}{\hbar^2v_F^2}.
	\end{equation}
	
For a constant magnetic field, orthogonal to the graphene surface and pointing in the positive $z$ direction ($\vec{B}=B_0\hat{e}_z$ with $B_0>0$), the vector potential is selected as $\vec{A}=B_0x\hat{e}_y$. Introducing now the constant $\omega$ as:
	\begin{equation*}
		B_0=\frac{c\hbar}{2e}\omega \quad \rightarrow \quad \omega = \frac{2eB_0}{c\hbar},
	\end{equation*}
whose dimensions are $($lenght$)^{-2}$, the potentials in (\ref{5}) become two shifted oscillators of the form
	\begin{equation}\label{6}
		V^\pm=\frac{\omega^2}{4}\left(x+\frac{2k}{\omega}\right)^2\pm\frac{1}{2}\omega.
	\end{equation}
	Thus, the eigenvalues $\mathcal{E}^\pm_n$ for the Hamiltonians $H^\pm$ are related as follows:
	\begin{equation}\label{7}
		\mathcal{E}^{-}_0=0, \quad \mathcal{E}^{-}_n=\mathcal{E}^{+}_{n-1}=n\omega, \quad n=1,2,\dots
	\end{equation}
and the associated eigenfunctions are those of the standard harmonic oscillator:
	\begin{equation}\label{8}
		\psi^{\pm}_n(x)=\sqrt{\frac{1}{2^nn!}\left(\frac{\omega}{2\pi}\right)^{1/2}}H_n\left[\sqrt{\frac{\omega}{2}}\left(x+\frac{2k}{\omega}\right)\right]e^{-\frac{\omega}{4}\left(x+\frac{2k}{\omega}\right)^2},
	\end{equation}
where $H_n$ denotes the Hermite polynomial of degree $n\in \mathbb{N}$.
	
We conclude that the complete solution of the corresponding Dirac-Weyl equation in a constant magnetic field consists of the eigenvalues
	\begin{equation}\label{9}
		E^\pm_n=\pm\hbar\,v_F\sqrt{n\omega}, \quad n=0,1,\ldots,
	\end{equation}
where the plus (minus) sign refers to the enegy electrones (holes), and the normalized eigenvectors
	\begin{equation}\label{10}
		\Psi_0(x,y)=e^{iky}\left(
		\begin{array}{c}
			0 \\
			i\,\psi^-_0(x) \\
		\end{array}
		\right), \quad \Psi_n(x,y)=\frac{e^{iky}}{\sqrt{2}}\left(
		\begin{array}{c}
			\psi^+_{n-1}(x) \\
			i\,\psi^-_n(x) \\
		\end{array}
		\right), \quad n=1,2,\ldots	
	\end{equation}
	
Note that the solutions of the Dirac-Weyl equation have an internal degree of freedom that mimics the spin, called pseudospin. It admits the following interpretation: each component is the projection of the particle wavefunction onto the sublattice A (spin up) or B (spin down).
		
	\section{\label{sec:level3}Annihilation operator}
	
Since the eigenstates of the previous Dirac-Weyl equation are expressed in terms of the eigenfunctions of the standard harmonic oscillator, it seems natural to look for an annihilation operator for the Hamiltonian in Eq.~(\ref{2}). In fact, let $\hat{A}^-$ be the operator defined by
	\begin{equation}\label{11}
		\hat{A}^-=\left(
		\begin{array}{cc}
			f_1(\hat{N})\hat{\vartheta}^- & 0 \\
			0 & f(\hat{N}+\hat{1})\hat{\vartheta}^- \\
		\end{array}
		\right),
	\end{equation}
where $\hat{\vartheta}^\pm$, $\hat{N}$ are given by
	\begin{equation*}
	\hat{\vartheta}^{-}=\frac{1}{\sqrt{2}}\left(z+\partial_z\right), \quad \hat{\vartheta}^{+}=\frac{1}{\sqrt{2}}\left(z-\partial_z\right), \quad \hat{N}=\hat{\vartheta}^{+}\hat{\vartheta}^-,
	\end{equation*}
with $z=\sqrt{\omega/2}\,(x+2k/\omega)$, and $f, \ f_1$ are two real adjustable functions which will be used to guarantee that $\hat{A}^-\Psi_n=c_n\Psi_{n-1}$. Then:
	\begin{eqnarray}\label{12}
		\hat{A}^-\Psi_n &=& 
		\begin{cases}
		0 & \text{for} \ n=0 \\
		\frac{f(1)}{\sqrt{2}} \Psi_0 & \text{for} \ n=1 \\
		\frac{e^{iky}}{\sqrt{2}}\left(
		\begin{array}{c}
			\sqrt{n-1}f_1(n-2)\psi_{n-2} \\
			\sqrt{n}f(n)i\psi_{n-1} \\
		\end{array}
		\right) & \text{for} \ n=2,3,\dots
		\end{cases}
	\end{eqnarray}
In order to ensure that
\begin{equation}\label{12.1}
\hat{A}^-\Psi_n=c_n\Psi_{n-1}
\end{equation}
it must happen that
	\begin{equation}\label{13}
		\sqrt{n-1}f_1(n-2)=\sqrt{n}f(n), \quad n=2,3,\dots
	\end{equation}
In such a case it is obtained that:
	\begin{eqnarray}\label{14}
		c_n &=& 
		\begin{cases}
		0 & \text{for} \ n=0 \\
		\frac{f(1)}{\sqrt{2}} & \text{for} \ n=1 \\
		\sqrt{n}f(n) & \text{for} \ n=2,3,\dots
		\end{cases}
	\end{eqnarray}
and the explicit expression for the annihilation operator $\hat{A}^-$ turns out to be:
    	 \begin{eqnarray}
\label{15} \hat{A}^-=\left(
    	 \begin{array}{cc}
    	 \frac{\sqrt{\hat{N}+\hat{2}}}{\sqrt{\hat{N}+\hat{1}}}f(\hat{N}+\hat{2})\hat{\vartheta}^- & 0 \\
    	 0 & f(\hat{N}+\hat{1})\hat{\vartheta}^-
    	 \end{array}\right).
    	 \end{eqnarray}
			
Equation (\ref{14}) indicates that the explicit form of the function $f(n)$ is required to determine the complete action of the annihilation operator $\hat{A}^-$ onto the eigenstates $\Psi_n$. It is also quite important for the properties of the graphene coherent states, as it will be immediately seen. 
			
\subsection{\textbf{Coherent states as eigenvectors of $\hat{A}^-$}}
	
Let us define the coherent states $\Psi_\alpha$ as eigenstates of the annihilation operator $\hat{A}^-$ with complex eigenvalue $\alpha$:
	\begin{equation}\label{16}
		\hat{A}^-\Psi_\alpha=\alpha\Psi_\alpha, \quad \alpha\in\mathbb{C}.
	\end{equation}
Expressing $\Psi_\alpha$ as a linear combination of the states $\Psi_n$, we have
	\begin{equation}\label{17}
		\Psi_\alpha(x,y)=\sum_{n=0}^\infty a_n\Psi_n(x,y)=\left[a_0 \left(\begin{array}{c}
			0 \\
			i\psi^-_0(x)
		\end{array}\right)+\sum_{n=1}^\infty \frac{a_n}{\sqrt{2}} \left(
		\begin{array}{c}
			\psi^+_{n-1}(x) \\
			i\psi^-_n(x) \\
		\end{array}
		\right)\right]e^{iky}.
	\end{equation}
Using equation (\ref{16}) we find a recurrence relation for the coefficients $a_n$ which depends on the value taken by $f(1)$. We can identify two different cases.
	
	\subsubsection{Case with $f(1)\neq0$}
	
First of all suppose that $f(n)\neq0 \ \forall \ n=1,2,\dots$ Thus we get $a_1={\sqrt{2}\alpha a_0}/{f(1)}$ and
	\begin{equation}\label{18}
	a_{n+1}=\frac{\alpha^n\,a_1}{\sqrt{(n+1)!}f(n+1)\cdots f(2)}=\frac{\sqrt{2}\,\alpha^{n+1}\,a_0}{\sqrt{(n+1)!}\,[f(n+1)]!},
	\end{equation}
where
	$$
  [f(n)]!\equiv \begin{cases}
	1 & \text{for} \  n=0 \\
	f(1)\cdots f(n) & \text{for} \ n=1,2,\dots
	\end{cases}
	$$
The free constant $a_0$ is used to normalize $\Psi_{\alpha}$; we obtain:
	\begin{equation}\label{21}
		\Psi_{\alpha}(x,y)=\left[1+\sum_{n=1}^\infty\frac{2\,|\alpha|^{2n}}{n!\,([f(n)]!)^2}\right]^{-1/2}\left[ 
		\Psi_{0}(x,y) + \sum_{n=1}^\infty \frac{\sqrt{2}\,\alpha^{n}}{\sqrt{n!}\,[f(n)]!} 
		\Psi_{n}(x,y)\right].
	\end{equation}
	
	\subsubsection{Case with $f(1)=0$}
If $f(1)=0$ we obtain that $a_0=0$ and the following recurrence relationship:
	\begin{equation}\label{22}
		a_{n+1}\sqrt{n+1}\,f(n+1)=\alpha \, a_n, \quad n=1,2,\dots
	\end{equation}
Now, depending on the value of $f(2)$, two possibilities appear once again.
	
	\paragraph*{\textbf{A. Case with $f(2)\neq0$.}}
If we suppose that $f(n)\neq0$ $\forall$ $n=2,3,\ldots$ and define $g(n)\equiv f(n+1)$, equation (\ref{22}) leads to:
	\begin{equation}\label{23}
		a_{n+1} = \frac{\alpha^n}{\sqrt{(n+1)!}[g(n)]!}a_1.
	\end{equation}
	Substituting this expression in equation (\ref{17}) and then normalizing we obtain:
	\begin{equation}\label{24}
		% \nonumber to remove numbering (before each equation)
		\Psi_\alpha(x,y) =  \left[\sum_{n=0}^\infty\frac{|\alpha|^{2n}}{(n+1)!\,([g(n)]!)^2}\right]^{-1/2}\sum_{n=0}^\infty \frac{\alpha^n}{\sqrt{(n+1)!}\,[g(n)]!}\Psi_{n+1}(x,y).
	\end{equation}
	
	\paragraph*{\textbf{B. Case with $f(2)=0$.}}
On the other hand, if $f(2)=0$ and $f(n)\neq0$ $\forall \ n=3,4,\dots$ the normalized coherent states turn out to be now:
	\begin{equation}\label{25}
		\Psi_\alpha(x,y) = \left[\sum_{n=0}^\infty\frac{|\alpha|^{2n}}{(n+2)!\,([h(n)]!)^2}\right]^{-1/2}\sum_{n=0}^\infty \frac{\alpha^n}{\sqrt{(n+2)!}\,[h(n)]!} \Psi_{n+2}(x,y),
	\end{equation}
where $h(n)\equiv f(n+2)$.

Let us notice that the graphene coherent states of Eqs. (\ref{21}, \ref{24}, \ref{25}) look similar to the so-called vector coherent states. For information concerning the last states, the reader can seek e.g. \cite{aeg04,ab08,ba09} and references therein.

\subsection{\textbf{Mean values and Heisenberg uncertainty relation}}
Let the dimensionless position and momentum operators be given by:
	\begin{equation}\label{26}
		\hat{z}=\frac{1}{\sqrt{2}}(\hat{\vartheta}^++\hat{\vartheta}^-), \quad \hat{p}=\frac{i}{\sqrt{2}}(\hat{\vartheta}^+-\hat{\vartheta}^-).
	\end{equation}
In units of $\hbar$, the Heisenberg uncertainty relation is expressed by
	\begin{equation}\label{27}
		\sigma_z^2\sigma_p^2\geq\frac{1}{4},
	\end{equation}
where $\sigma_S^2 \equiv\langle\hat{S}^2\rangle-\langle\hat{S}\rangle^2$ for an arbitrary observable $\hat{S}$.

We will calculate next these quantities for some examples of the coherent states, which will stress the important role played by the function $f(n)$ in our treatment.

\subsubsection{The case with $f(1)\neq0$.}
	
Let us consider first the particular choice $f(\hat{N})=\hat{1}$. Thus, expression (\ref{21}) leads to:
	\begin{equation}\label{28}
		\Psi_\alpha(x,y)=\frac{1}{\sqrt{2e^{r^2}-1}}\left[ \Psi_0(x,y)+\sum_{n=1}^\infty \frac{\sqrt{2}\alpha^{n}}{\sqrt{n!}} \Psi_n(x,y)\right],
	\end{equation}
where $r=|\alpha|$.
	
\begin{figure}
	\begin{minipage}[h]{0.5\textwidth}
		\includegraphics[width=\textwidth]{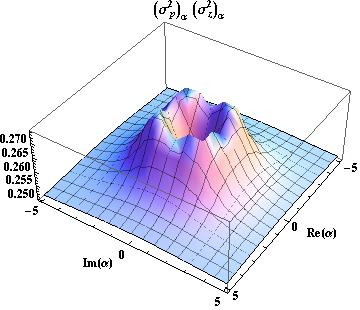}
	\end{minipage}
	\hfill
	\begin{minipage}[h]{0.45\textwidth}
		\includegraphics[width=\textwidth]{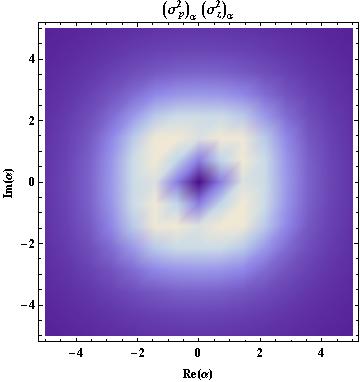}
	\end{minipage}
	\caption{Heisenberg uncertainty relation $(\sigma_z)^2_\alpha(\sigma_p)^2_\alpha$ as function of $\alpha$ for $f(n)=1$.}\label{fig:Graphene10}
\end{figure}

Using these coherent states, the mean values for the operators $\hat{z}, \hat{p}$ of Eq. (\ref{26}) as well as their squares become:
	\begin{subequations}
		\begin{align}\label{29}
			% \nonumber to remove numbering (before each equation)
			\langle\hat{z}\rangle_{\alpha} &= \frac{\sqrt{2}\text{Re}(\alpha)}{2e^{r^2}-1}\left[e^{r^2}+\sum_{n=1}^\infty\frac{r^{2n}}{\Gamma(n)\Gamma(n+2)}\right], \\
			\langle\hat{z}^2\rangle_{\alpha} &= \frac{1}{4e^{r^2}-2}\left[1+4r^2e^{r^2}+2[[\text{Re}(\alpha)]^2-[\text{Im}(\alpha)]^2]\left(e^{r^2}+\sum_{n=1}^\infty\frac{\sqrt{n+1}r^{2n}}{\Gamma(n)\Gamma(n+3)}\right)\right], \\
			\langle\hat{p}\rangle_{\alpha} &= \frac{\sqrt{2}\text{Im}(\alpha)}{2e^{r^2}-1}\left[e^{r^2}+\sum_{n=1}^\infty\frac{r^{2n}}{\Gamma(n)\Gamma(n+2)}\right], \\
		\langle\hat{p}^2\rangle_{\alpha} &= \frac{1}{4e^{r^2}-2}\left[1+4r^2e^{r^2}-2[[\text{Re}(\alpha)]^2-[\text{Im}(\alpha)]^2]\left(e^{r^2}+\sum_{n=1}^\infty\frac{\sqrt{n+1}r^{2n}}{\Gamma(n)\Gamma(n+3)}\right)\right].
		\end{align}
	\end{subequations}
Through them it is straightforward to calculate $(\sigma_z)^2_\alpha(\sigma_p)^2_\alpha$ (see Figure \ref{fig:Graphene10}). Note that in the limit $\alpha\rightarrow0$ we have $(\sigma_z)^2_\alpha(\sigma_p)^2_\alpha\rightarrow1/4$. %\ref{fig:Graphene11}).
	\subsubsection{The case with $f(1)=0$.}
As we saw in section 3.1.2, when $f(1)=0$ two options appear, which depend on the value taken by $f(2)$.
	
	\paragraph*{\textbf{A. The case with $f(2)\neq0$.}}
Let us choose now $f(\hat{N}+\hat{1})=g(\hat{N})=\frac{\sqrt{\hat{N}}}{\sqrt{\hat{N}+\hat{1}}}$, so that $f(n)\neq0$ $\forall$ $n=2,3,\ldots$. From Eq.~(\ref{24}), the explicit form for the normalized coherent states becomes:
	\begin{equation}\label{29.1}
	\Psi_\alpha(x,y)=e^{-r^2/2}\sum_{n=0}^\infty \frac{\alpha^n}{\sqrt{n!}}\Psi_{n+1}(x,y).
	\end{equation}
	
\begin{figure}
					\begin{minipage}[h]{0.5\textwidth}
						\includegraphics[width=\textwidth]{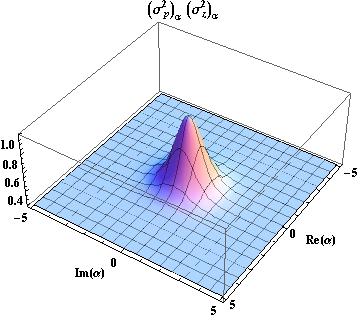}
					\end{minipage}
					\hfill
					\begin{minipage}[h]{0.45\textwidth}
						\includegraphics[width=\textwidth]{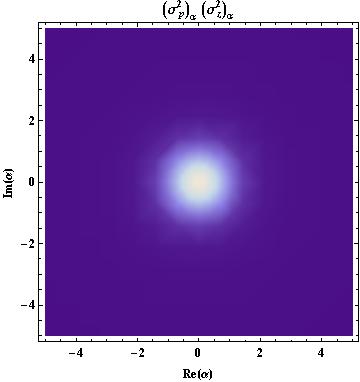}
					\end{minipage}
					\caption{Heisenberg uncertainty relation $(\sigma_z)^2_\alpha(\sigma_p)^2_\alpha$ as function of $\alpha$ for $f(n)=\sqrt{n-1}/\sqrt{n}$.}\label{fig:Graphene01}
				\end{figure}

The mean values for the operators of Eq. (\ref{26}) and their squares become:
	\small
	%\scriptsize
	\begin{subequations}
		\begin{align}\label{29.2}
		% \nonumber to remove numbering (before each equation)
		\langle\hat{z}\rangle_{\alpha} &= \frac{\text{Re}(\alpha)}{\sqrt{2}}\left[1+e^{-r^2}\sum_{n=0}^\infty\frac{\sqrt{n+2}\,r^{2n}}{\sqrt{\Gamma(n+1)\Gamma(n+2)}}\right], \\
		\langle\hat{z}^2\rangle_{\alpha} &= e^{-r^2}\sum_{n=0}^\infty\frac{(n+1)r^{2n}}{\Gamma(n+1)}+\frac{[[\text{Re}(\alpha)]^2-[\text{Im}(\alpha)]^2]}{2}\left(1+e^{-r^2}\sum_{n=0}^\infty\frac{\sqrt{n+3}\,r^{2n}}{\sqrt{\Gamma(n+1)\Gamma(n+2)}}\right), \\
		\langle\hat{p}\rangle_{\alpha} &= \frac{\text{Im}(\alpha)}{\sqrt{2}}\left[1+e^{-r^2}\sum_{n=0}^\infty\frac{\sqrt{n+2}\,r^{2n}}{\sqrt{\Gamma(n+1)\Gamma(n+2)}}\right], \\
		\langle\hat{p}^2\rangle_{\alpha} &= e^{-r^2}\sum_{n=0}^\infty\frac{(n+1)r^{2n}}{\Gamma(n+1)}-\frac{[[\text{Re}(\alpha)]^2-[\text{Im}(\alpha)]^2]}{2}\left(1+e^{-r^2}\sum_{n=0}^\infty\frac{\sqrt{n+3}\,r^{2n}}{\sqrt{\Gamma(n+1)\Gamma(n+2)}}\right).
		\end{align}
	\end{subequations}
	\normalsize
In the limit $\alpha\rightarrow0$ it turns out that $(\sigma_z)^2_\alpha(\sigma_p)^2_\alpha\rightarrow1$ (see Figure \ref{fig:Graphene01}).

	\paragraph*{\textbf{B. The case with $f(2)=0$.}}
Let us consider finally that $f(\hat{N}+\hat{2})=h(\hat{N})=\frac{\hat{N}\sqrt{\hat{N}+\hat{1}}}{\sqrt{\hat{N}+\hat{2}}}$. The explicit expression for the normalized coherent states arises from equation (\ref{25}):
	\begin{equation}\label{32}
		\Psi_\alpha(x,y)=\frac{1}{\sqrt{_0F_2(1,2;r^2)}}\sum_{n=0}^\infty \frac{\alpha^n}{n!\sqrt{(n+1)!}}\Psi_{n+2}(x,y),
	\end{equation}
where $_pF_q$ is a generalized hypergeometric function defined by
	\begin{equation*}
_pF_q(a_1,\ldots,a_p,b_1,\ldots,b_q;x)=\frac{\Gamma(b_1)\ldots\Gamma(b_q)}{\Gamma(a_1)\ldots\Gamma(a_p)}\sum_{n=0}^{\infty}\frac{\Gamma(a_1+n)\ldots\Gamma(a_p+n)}{\Gamma(b_1+n)\ldots\Gamma(b_q+n)}\frac{x^n}{n!}.
	\end{equation*}
%				\begin{figure}
%					\begin{minipage}[h]{0.5\textwidth}
%						\includegraphics[width=\textwidth]{Graphene21}
%					\end{minipage}
%					\hfill
%					\begin{minipage}[h]{0.45\textwidth}
%						\includegraphics[width=\textwidth]{Graphene211}
%					\end{minipage}
%					\caption{Uncertainty relation $(\sigma^2_z)_\alpha(\sigma^2_p)_\alpha$ as a function of $\alpha$. The behavior for the case $f_2(0)=0$ and $f(0)\neq0$ is shown.}
%				\end{figure}\label{fig:Graphene21}

\begin{figure}
					\begin{minipage}[h]{0.5\textwidth}
						\includegraphics[width=\textwidth]{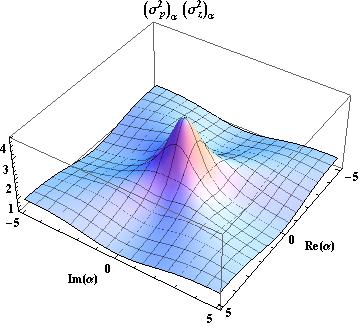}
					\end{minipage}
					\hfill
					\begin{minipage}[h]{0.45\textwidth}
						\includegraphics[width=\textwidth]{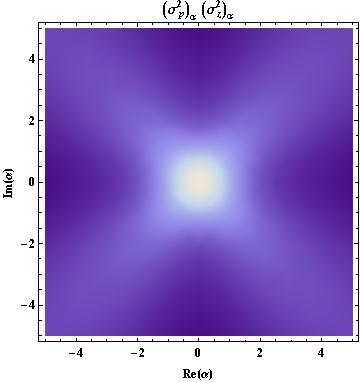}
					\end{minipage}
					\caption{Heisenberg uncertainty relation $(\sigma_z)^2_\alpha(\sigma_p)^2_\alpha$ as function of $\alpha$ for $f(n)=(n-2)\sqrt{n-1}/\sqrt{n}$.}\label{fig:Graphene31}
				\end{figure}

The mean values for the operators in Eq. (\ref{26}) and their squares are now:
	\small
	%\footnotesize
	%\scriptsize
	%\tiny
	\begin{subequations}
		\begin{align}\label{33}
			% \nonumber to remove numbering (before each equation)
			\langle\hat{z}\rangle_{\alpha} &= \frac{\text{Re}(\alpha)}{\sqrt{2}\,_0F_2(1,2;r^2)}\left[\,_0F_2(2,2;r^2)+\sum_{n=0}^\infty\frac{\sqrt{n+3}\,r^{2n}}{n!\sqrt{[\Gamma(n+2)]^3\Gamma(n+3)}}\right], \\
			\nonumber  \langle\hat{z}^2\rangle_{\alpha} &= \frac{1}{2\,_0F_2(1,2;r^2)}\left[2\sum_{n=0}^\infty\frac{(n+2)r^{2n}}{\Gamma(n+2)[\Gamma(n+1)]^{2}}+[[\text{Re}(\alpha)]^2-[\text{Im}(\alpha)]^2]\times\right. \\
			&\quad \left.\times\left(\frac{_0F_2(2,3;r^2)}{2}+\sum_{n=0}^\infty\frac{\sqrt{n+3}\,r^{2n}}{\Gamma(n+1)\sqrt{\Gamma(n+2)[\Gamma(n+3)]^3}}\right)\right], \\
			\langle\hat{p}\rangle_{\alpha} &= \frac{\text{Im}(\alpha)}{\sqrt{2}\,_0F_2(1,2;r^2)}\left[\,_0F_2(2,2;r^2)+\sum_{n=0}^\infty\frac{\sqrt{n+3}\,r^{2n}}{n!\sqrt{[\Gamma(n+2)]^3\Gamma(n+3)}}\right], \\
			\nonumber  \langle\hat{p}^2\rangle_{\alpha} &= \frac{1}{2\,_0F_2(1,2;r^2)}\left[2\sum_{n=0}^\infty\frac{(n+2)r^{2n}}{\Gamma(n+2)[\Gamma(n+1)]^{2}}-[[\text{Re}(\alpha)]^2-[\text{Im}(\alpha)]^2]\times\right. \\
			&\quad  \left.\times\left(\frac{_0F_2(2,3;r^2)}{2}+\sum_{n=0}^\infty\frac{\sqrt{n+3}\,r^{2n}}{\Gamma(n+1)\sqrt{\Gamma(n+2)[\Gamma(n+3)]^3}}\right)\right].
		\end{align}
	\end{subequations}
	\normalsize
In the limit $\alpha\rightarrow0$ we get $(\sigma_z)^2_\alpha(\sigma_p)^2_\alpha\rightarrow4$ (see Figure \ref{fig:Graphene31}).
\begin{figure}
	\begin{minipage}[h]{0.32\textwidth}
		\includegraphics[width=\textwidth]{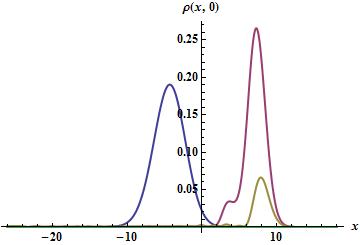}%{GrapheneDensity03D}
	\end{minipage}
	%	\hfill
	\hspace{0.1cm}
	\begin{minipage}[h]{0.3\textwidth}
		\includegraphics[width=\textwidth]{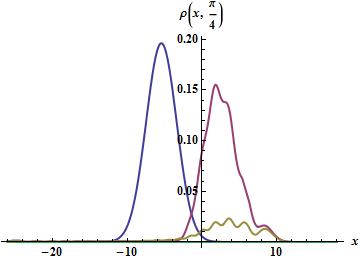}%{GrapheneDensity0}
	\end{minipage}
	%	\hfill
	\hspace{0.1cm}
	\begin{minipage}[h]{0.3\textwidth}
		\includegraphics[width=\textwidth]{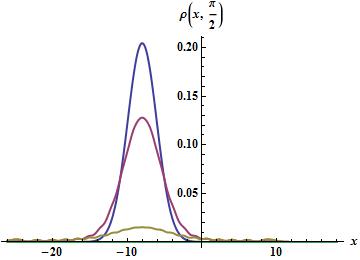}%{GrapheneDensity0}
	\end{minipage}
%					\end{figure}
%					\begin{figure}
	\begin{minipage}[h]{0.3\textwidth}
		\includegraphics[width=\textwidth]{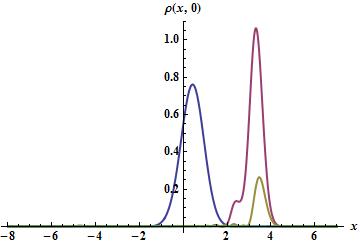}%{GrapheneCurrent03D}
	\end{minipage}
	%\hfill
	\hspace{0.55cm}
	\begin{minipage}[h]{0.3\textwidth}
		\includegraphics[width=\textwidth]{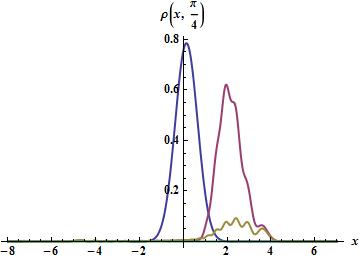}%{GrapheneCurrent0-1}
	\end{minipage}
	%\hfill
	\hspace{0.55cm}
	\begin{minipage}[h]{0.3\textwidth}
		\includegraphics[width=\textwidth]{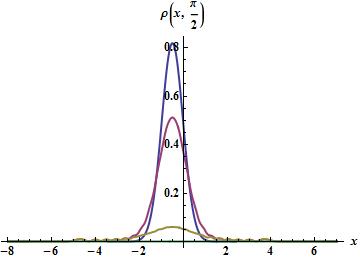}%{GrapheneCurrent0-1}
	\end{minipage}
		\caption{Probability density $\rho(x,r,\theta)$ for $f(n)=1$ with $B_0=1/8$ (above), $B_0=2$ (below), $\theta=0, \pi/4, \pi/2$ (left to right respectively) and $k=1$. The blue, red and brown lines correspond to $r=1,4,5$ respectively.}\label{Den1}%For the case $f_2(0)\neq0$, with a constant magnetic $B(x)=1/2$ and $\omega=k=1$, the probability density $\rho(x)$ is shown above; the current density $j(x)/(ev_F)$ is shown below. Both functions are plotted for $\theta=0, \pi/4, \pi/2$. Blue line corresponds to $r=1$; red line to $r=10$, and brown line to $r=20$.}
\end{figure}%\label{Cur1}
				
As we can see, the Heisenberg uncertainty relation depends strongly on the coherent states under consideration. Thus, for the states in Eq.~(\ref{28}) it takes its minimum at $\alpha=0$, while for those in Eqs.~(\ref{29.1}) and (\ref{32}) their maxima are reached at the same point; this is so since the lowest energy eigenstate involved in the corresponding linear combination is different if different families of coherent states are taken into account (see also \cite{fhn94,fnr95,fh99}).
		
%\newpage
\subsection{Magnetic field and probability density}%and current densities.}
\begin{figure}
	\begin{minipage}[h]{0.32\textwidth}
		\includegraphics[width=\textwidth]{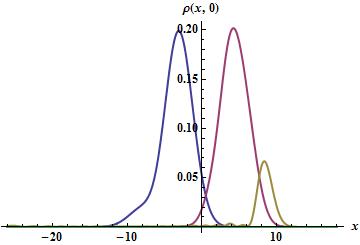}%{GrapheneDensity03D}
	\end{minipage}
%	\hfill
\hspace{0.1cm}
	\begin{minipage}[h]{0.3\textwidth}
		\includegraphics[width=\textwidth]{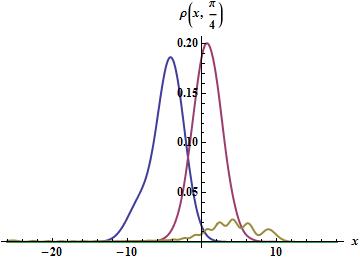}%{GrapheneDensity0}
	\end{minipage}
%	\hfill
\hspace{0.1cm}
	\begin{minipage}[h]{0.3\textwidth}
		\includegraphics[width=\textwidth]{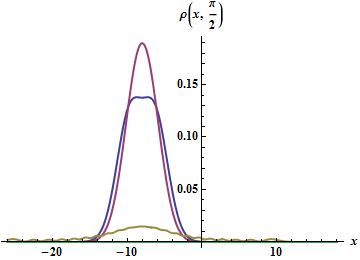}%{GrapheneDensity0}
	\end{minipage}
%	\end{figure}
%	\begin{figure}
	\begin{minipage}[h]{0.3\textwidth}
		\includegraphics[width=\textwidth]{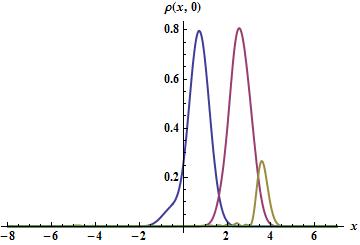}%{GrapheneCurrent03D}
	\end{minipage}
	%\hfill
	\hspace{0.55cm}
	\begin{minipage}[h]{0.3\textwidth}
		\includegraphics[width=\textwidth]{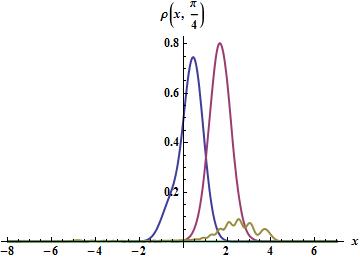}%{GrapheneCurrent0-1}
	\end{minipage}
		%\hfill
		\hspace{0.55cm}
		\begin{minipage}[h]{0.3\textwidth}
			\includegraphics[width=\textwidth]{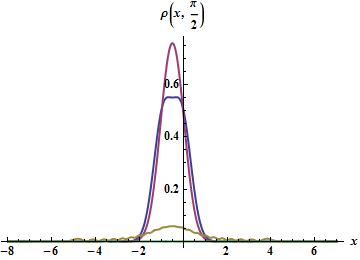}%{GrapheneCurrent0-1}
		\end{minipage}
	\caption{Probability density $\rho(x,r,\theta)$ for $f(n)=\sqrt{n-1}/\sqrt{n}$ with $B_0=1/8$ (above), $B_0=2$ (below), $\theta=0, \pi/4, \pi/2$ (left to right respectively) and $k=1$. The blue, red and brown lines correspond to $r=1,3,5$ respectively.}\label{Den0}%For the case $f(0)\neq0$, with a constant magnetic $B(x)=1/2$ and $\omega=k=1$, the probability density $\rho(x)$ is shown above; the current density $j(x)/(ev_F)$ is shown below. Both functions are plotted for $\theta=0, \pi/4, \pi/2$. Blue line corresponds to $r=1$; red line to $r=3$, and brown line to $r=5$.}
\end{figure}%\label{Cur0}
 The probability density $\rho = \Psi_\alpha^\dagger \Psi_\alpha$ will be used to analyze the properties of the graphene coherent states. It will depend on the following matrix elements \cite{kuru}:
%\begin{subequations}
	\begin{eqnarray}
	% \nonumber to remove numbering (before each equation)
	\rho_{n,m}(x) &:=& \psi^{+}_{n-1}(x)\psi^+_{m-1}(x)+\psi^-_n(x)\psi^-_m(x)=\rho_{m,n}(x). \label{34}
%	\\
%	j_{n,m}(x) &:=& 2ev_F\psi_{n-1}(x)\psi_m(x), \label{34b}
	\end{eqnarray}
%\end{subequations}
Note also that, according to Eq.~(\ref{8}), it depends on the magnetic field intensity $B_0$ through the parameter $\omega$.
\subsubsection{Probability density for $f(1)\neq0$.}
A straightforward calculation using Eq.~(\ref{28}) leads to:
%\small
%\footnotesize
%\begin{subequations}
	\begin{align}\label{35}
	% \nonumber to remove numbering (before each equation)
	\rho(x,r,\theta) &= \frac{1}{2e^{r^2}-1}\left[\sum_{m=1}^\infty\sum_{n=1}^\infty\frac{r^{m+n}\cos{(n-m)\theta}}{\sqrt{m!\,n!}}\rho_{n,m}(x)\right. \nonumber\\
	&\quad \left.+2\sum_{n=1}^\infty\frac{r^n\cos(n\theta)}{\sqrt{n!}}\psi^-_n(x)\psi^-_0(x)+(\psi^-_0(x))^2\right],
	\end{align}
%\end{subequations}
%\normalsize
where $\theta=$Arg$[\alpha]$, $r=\vert\alpha\vert$. Plots of this probability density for two magnetic field intensities and different $\theta$'s and $r$'s are shown in Figure \ref{Den1}.
%\begin{figure}
%	\begin{minipage}[h]{0.32\textwidth}
%		\includegraphics[width=\textwidth]{Den2_0}%{GrapheneDensity03D}
%	\end{minipage}
%	%	\hfill
%	\hspace{0.1cm}
%	\begin{minipage}[h]{0.3\textwidth}
%		\includegraphics[width=\textwidth]{Den2_pi4}%{GrapheneDensity0}
%	\end{minipage}
%	%	\hfill
%	\hspace{0.1cm}
%	\begin{minipage}[h]{0.3\textwidth}
%		\includegraphics[width=\textwidth]{Den2_pi2}%{GrapheneDensity0}
%	\end{minipage}
%	\end{figure}
%	\label{Den2}
%	\begin{figure}
%	\begin{minipage}[h]{0.32\textwidth}
%		\includegraphics[width=\textwidth]{Cur2_0}%{GrapheneDensity03D}
%	\end{minipage}
%	%	\hfill
%	\hspace{0.1cm}
%	\begin{minipage}[h]{0.3\textwidth}
%		\includegraphics[width=\textwidth]{Cur2_pi4}%{GrapheneDensity0}
%	\end{minipage}
%	%	\hfill
%	\hspace{0.1cm}
%	\begin{minipage}[h]{0.3\textwidth}
%		\includegraphics[width=\textwidth]{Cur2_pi2}%{GrapheneDensity0}
%	\end{minipage}
%	\caption{For the case $f(0)\neq0$, with a constant magnetic $B(x)=1/2$ and $\omega=k=1$, the probability density $\rho(x)$ is shown above; the current density $j(x)/(ev_F)$ is shown below. Both functions are plotted for $\theta=0, \pi/4, \pi/2$. Blue line corresponds to $r=1$; red line to $r=50$, and brown line to $r=100$.}
%\end{figure}\label{Cur2}
\subsubsection{Probability density for $f(1)=0$}
\paragraph*{\textbf{A. Case with $f(2)\neq0$.}}
%Image
On the other hand, for the states of Eq. (\ref{29.1}) we get (see Figure \ref{Den0}):
%\small
%\footnotesize
%\begin{subequations}
	\begin{align}\label{35.1}
	% \nonumber to remove numbering (before each equation)
	\rho(x,r,\theta) &= \frac{e^{-r^2}}{2} \sum_{m=0}^\infty\sum_{n=0}^\infty\frac{r^{m+n}\cos{(n-m)\theta}}{\sqrt{\Gamma(m+1)\Gamma(n+1)}}\rho_{n+1,m+1}(x). %\\
%	j(x,r,\theta) &=   \frac{ev_Fe^{-r^2}}{2} \nonumber \\
%	&\quad \sum_{m=0}^\infty\sum_{n=0}^\infty\frac{r^{m+n}\cos{(n-m)\theta}}{\sqrt{\Gamma(m+1)\Gamma(n+1)}}\frac{j_{m+1,n+1}(x)+j_{n+1,m+1}(x)}{2ev_F}.
	\end{align}
%\end{subequations}
%\normalsize
   \begin{figure}
   	\begin{minipage}[h]{0.3\textwidth}
   		\includegraphics[width=\textwidth]{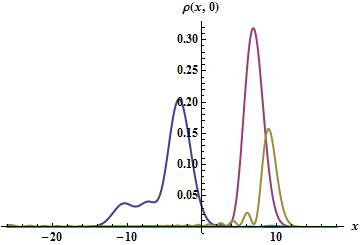}%{GrapheneDensity33D}
   	\end{minipage}
   	%\hfill
   	\hspace{0.1cm}
   	\begin{minipage}[h]{0.3\textwidth}
   		\includegraphics[width=\textwidth]{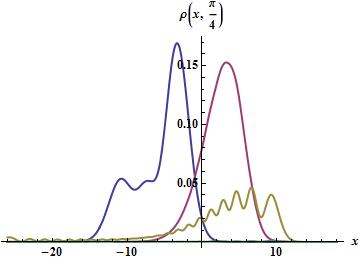}%{GrapheneDensity3}
   	\end{minipage}
   	   	\hspace{0.1cm}
   	   	\begin{minipage}[h]{0.3\textwidth}
   	   		\includegraphics[width=\textwidth]{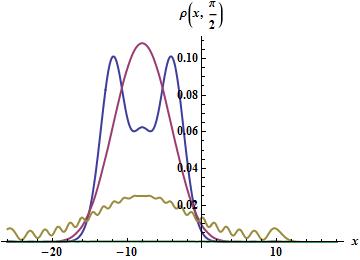}%{GrapheneDensity3}
   	   	\end{minipage}
 %   \end{figure}
%   	\begin{figure}
   	\begin{minipage}[h]{0.3\textwidth}
   		\includegraphics[width=\textwidth]{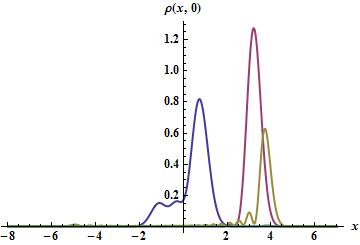}%{GrapheneCurrent33D}
   	\end{minipage}
   	%\hfill
   	\hspace{0.55cm}
   	\begin{minipage}[h]{0.3\textwidth}
   		\includegraphics[width=\textwidth]{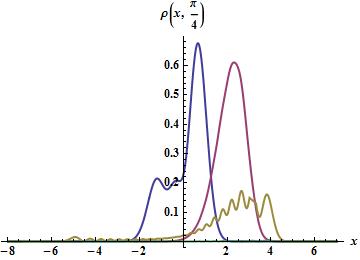}%{GrapheneCurrent3}
   	\end{minipage}
   	   	\hspace{0.55cm}
   	   	\begin{minipage}[h]{0.3\textwidth}
   	   		\includegraphics[width=\textwidth]{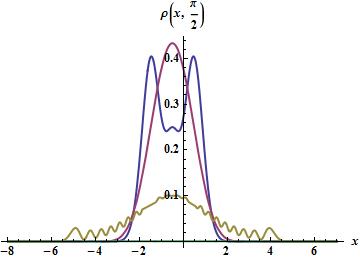}%{GrapheneCurrent3}
   	   	\end{minipage}
   	\caption{Probability density $\rho(x,r,\theta)$ for $f(n)=(n-2)\sqrt{n-1}/\sqrt{n}$ with $B_0=1/8$ (above), $B_0=2$ (below), $\theta=0, \pi/4, \pi/2$ (left to right respectively), and $k=1$. The blue, red and brown lines correspond to $r=1,50,100$ respectively.}\label{Den3}%For the case $f(0)=0$, with a constant magnetic $B(x)=1/2$ and $\omega=k=1$, the probability density $\rho(x)$ is shown above; the current density $j(x)/(ev_F)$ is shown below. Both functions are plotted for $\theta=0, \pi/4, \pi/2$. Blue line corresponds to $r=1$; red line to $r=50$, and brown line to $r=100$.}
   \end{figure}%\label{Cur3}
%Now, for the states in Eq. (\ref{30}), we have (Figure \ref{Den2})
%%\small
%%\footnotesize
%\begin{subequations}
%	\begin{align}\label{36}
%	% \nonumber to remove numbering (before each equation)
%	\rho(x,r,\theta) &= \frac{1}{2\,_0F_2(1,1;r^2)} \sum_{m=0}^\infty\sum_{n=0}^\infty\frac{r^{m+n}\cos{(n-m)\theta}}{\sqrt{[\Gamma(m+1)\Gamma(n+1)]^3}}\rho_{n+1,m+1}(x), \\
%	j(x,r,\theta) &=   \frac{ev_F}{2\,_0F_2(1,1;r^2)} \nonumber \\
%	&\quad \sum_{m=0}^\infty\sum_{n=0}^\infty\frac{r^{m+n}\cos{(n-m)\theta}}{\sqrt{[\Gamma(m+1)\Gamma(n+1)}]^3}\frac{j_{m+1,n+1}(x)+j_{n+1,m+1}(x)}{2ev_F}.
%	\end{align}
%\end{subequations}
%%\normalsize
\paragraph*{\textbf{B. Case with $f(2)=0$.}}
%Image
Finally, by employing the states of Eq.~(\ref{32}) we arrive at (see Figure \ref{Den3})
%\footnotesize 
%\begin{subequations}
	\begin{align}\label{37}
	% \nonumber to remove numbering (before each equation)
	\rho(x,r,\theta) &= \frac{1}{2\,_0F_2(1,2;r^2)} \sum_{m=0}^\infty\sum_{n=0}^\infty\frac{r^{m+n}\cos{(n-m)\theta}}{m!\,n!\sqrt{\Gamma(m+2)\Gamma(n+2)}}\rho_{n+2,m+2}(x).%, \\
%	j(x,r,\theta) &= \frac{ev_F}{2\,_0F_2(1,2;r^2)} \nonumber \\
%	&\quad\sum_{m=0}^\infty\sum_{n=0}^\infty\frac{r^{m+n}\cos{(n-m)\theta}}{\Gamma(m+1)\Gamma(n+1)\sqrt{\Gamma(m+2)\Gamma(n+2)}}\frac{j_{m+2,n+2}(x)+j_{n+2,m+2}(x)}{2ev_F}.
	\end{align}
%\end{subequations}

\subsubsection{Discussion}
For the graphene coherent states under study, the probability density reaches a maximum which displaces along the $x$ direction as the parameter $\theta\in[0,\pi/2]$ grows. Also, when $r\rightarrow\infty$ we have %both 
   $\rho(r)\rightarrow0$ %and $j(r)\rightarrow0$, 
   $\forall\,\theta$.
   
The previous behavior can be interpreted as follows. According to the mean values of the position and momentum operators, they can be expressed in terms of the complex eigenvalue $\alpha$ in the way: 
	$$
	\langle\hat{z}\rangle\sim\text{Re}(\alpha)F_1(\vert\alpha\vert),\quad \langle\hat{p}\rangle\sim\text{Im}(\alpha)F_2(\vert\alpha\vert),
	$$ 
where $F_{1,2}(\vert\alpha\vert)$ are certain functions that depend on the graphene coherent states under study. In particular, when $\alpha$ is real (for $\theta=0$) we have $\langle\hat{p}\rangle=0$ which means that, on average, the electron \textit{moves} as many times to the right as to the left, canceling out at the end the positive momentum contributions with the negative ones. Meanwhile, when $\alpha$ is purely imaginary (for $\theta=\pi/2$) we have that $\langle\hat{z}\rangle=0$. This can be interpreted as if the system would perform symmetric oscillations around the equilibrium position $z_0$ (or potential center), which is determined by the magnetic field intensity.
   
On the other hand, when $B_0$ increases the maximum of the probability density also grows up while their width decreases (due to probability conservation). This means that the electron is to be found in a more bounded region as $B_0$ grows. An opposite interpretation can be formulated when $B_0$ decreases.

  \subsection{Mean energy value}
The mean energy value, $\langle\hat{H}\rangle_\alpha$, is another quantity useful to characterize the graphene coherent states.
   
According to the expansion in Eq. (\ref{17}), where $\Psi_n$ are the eigenfunctions of the Dirac-Weyl Hamiltonian (see Eq.~(\ref{2})), for the graphene coherent states we have that
  \begin{equation}\label{37.1}
  \langle\hat{H}\rangle_\alpha=\sum_{n=0}^{\infty}E_n|a_n|^2=\hbar v_F\sum_{n=0}^{\infty}\sqrt{n\omega}|a_n|^2.
  \end{equation}

The mean energy value is calculated for each coherent state using this expression, which leads to the following results.
          	         	\begin{figure}
          	         		\centering
          	         		\begin{minipage}[h]{0.40\textwidth}
          	         			\includegraphics[width=\textwidth]{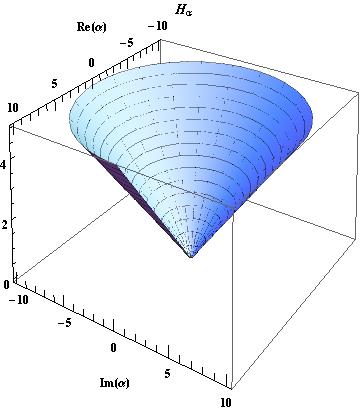}
          	         		\end{minipage}
          	         		\hspace{1.5cm}%\hfill
          	         		\begin{minipage}[h]{0.40\textwidth}
          	         			\includegraphics[width=\textwidth]{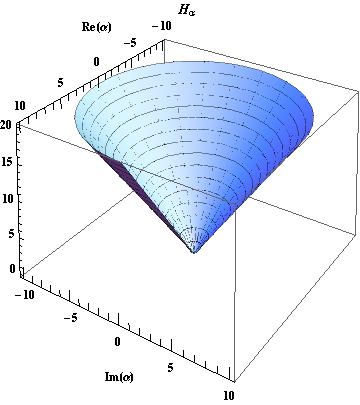}
          	         		\end{minipage}
          	         		\caption{Mean energy value of Eq.~(\ref{38}) for $f(n)=1$, with $k=1$ and the magnetic field intensities $B_0=1/8$ (left) and $B_0=2$ (right).}\label{Energy1} 
          	         		\end{figure}

          	         		\begin{figure}
          	         		\centering
          	         		\begin{minipage}[h]{0.40\textwidth}
          	         			\includegraphics[width=\textwidth]{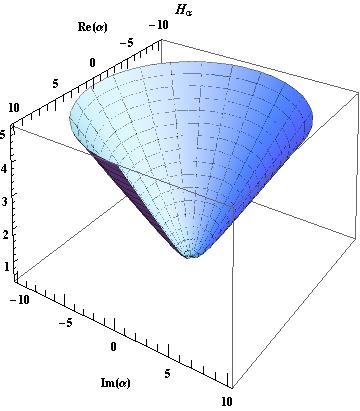}
          	         		\end{minipage}
          	         		\hspace{1.5cm}%\hfill
          	         		\begin{minipage}[h]{0.40\textwidth}
          	         			\includegraphics[width=\textwidth]{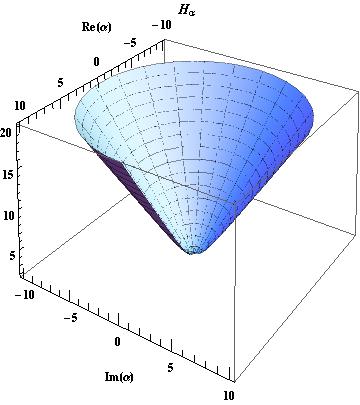}
          	         		\end{minipage}
          	         		\caption{Mean energy value of Eq.~(\ref{39}) for $f(n)=\sqrt{n-1}/\sqrt{n}$, with $k=1$ and the magnetic field intensities $B_0=1/8$ (left) and $B_0=2$ (right).}\label{Energy0}
          	         	\end{figure}
   \subsubsection{$\langle\hat{H}\rangle_\alpha$ for $f(1)\neq0$.}
For the coherent states of Eq.~(\ref{28}) we obtain (see Figure \ref{Energy1}):
   \begin{equation}\label{38}
   	\langle\hat{H}\rangle_{\alpha}=\frac{2\hbar v_F}{2e^{r^2}-1}\sum_{n=0}^\infty\frac{r^{2n}\sqrt{n\omega}}{\Gamma(n+1)}.
   \end{equation}
   \subsubsection{$\langle\hat{H}\rangle_\alpha$ for $f(1)=0$}
   \paragraph*{\textbf{A. Case with $f(2)\neq0$.}} For the coherent states of Eq.~(\ref{29.1}) we get (see Figure \ref{Energy0}):
   \begin{equation}\label{39}
   	\langle\hat{H}\rangle_{\alpha}=\hbar v_Fe^{-r^2}\sum_{n=0}^\infty\frac{r^{2n}\sqrt{(n+1)\omega}}{\Gamma(n+1)}.
   \end{equation}
   
\begin{figure}
                 		\centering
%                 		\begin{minipage}[h]{0.3\textwidth}
%                 			\includegraphics[width=\textwidth]{GrapheneEnergy23D}
%                 		\end{minipage}
%                 		\hspace{3cm}%\hfill
%                 		\begin{minipage}[h]{0.3\textwidth}
%                 			\includegraphics[width=\textwidth]{GrapheneEnergy213D}
%                 		\end{minipage}
%                 		\caption{For the case $f_2(0)=0$ and $f(0)\neq0$, with a constant magnetic $B(x)=1/2$ and $\omega=k=1$, the mean value of energy in Eq. (\ref{40}) is shown.}
%                 		%\end{figure}
%                 		\label{Energy2}
                 		%\begin{figure}
                 		\begin{minipage}[h]{0.4\textwidth}
                 			\includegraphics[width=\textwidth]{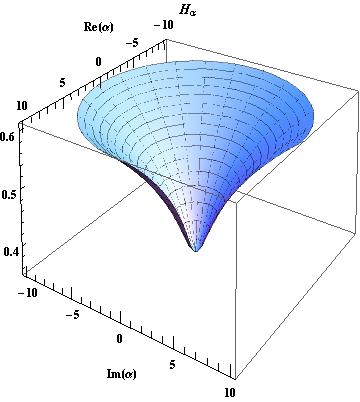}
                 		\end{minipage}
                 		\hspace{1.5cm}%\hfill
                 		\begin{minipage}[h]{0.4\textwidth}
                 			\includegraphics[width=\textwidth]{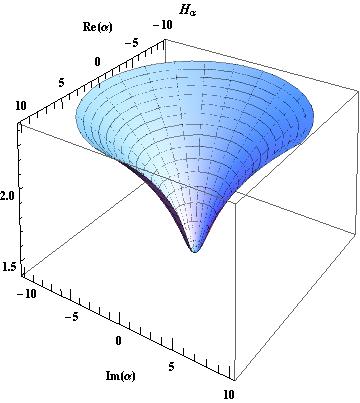}
                 		\end{minipage}
                 		\caption{Mean energy value of Eq.~(\ref{41}) for $f(n)=(n-2)\sqrt{n-1}/\sqrt{n}$ with $k=1$ and the magnetic field intensities $B_0=1/8$ (left) and $B_0=2$ (right).}\label{Energy3}
                 	\end{figure}

%      	Now, for the NLCS shown in Eq.~(\ref{30}), we have (Figure \ref{Energy2}):
%   \begin{equation}\label{40}
%   	\langle\hat{H}\rangle^{III}_\alpha=\frac{\hbar v_F}{_0F_2(1,1;r^2)}\sum_{n=0}^\infty\frac{r^{2n}\sqrt{(n+1)\omega}}{[\Gamma(n+1)]^3}.
%   \end{equation}
   \paragraph*{\textbf{B. Case with $f(2)=0$.}}
         	Finally, for the coherent states of Eq.~(\ref{32}) we arrive at (see Figure \ref{Energy3}):
         	\begin{equation}\label{41}
         	\langle\hat{H}\rangle_{\alpha}=\frac{\hbar v_F}{\,_0F_2(1,2;r^2)}\sum_{n=0}^\infty\frac{r^{2n}\sqrt{(n+2)\omega}}{\Gamma[n+2][\Gamma(n+1)]^2}.
         	\end{equation}
While Figures \ref{Energy1} and \ref{Energy0} have a similar qualitative behavior for $\langle\hat{H}\rangle_\alpha$ as function of $\alpha$, Figure \ref{Energy3} shows that the mean energy value for the states of Eq.~(\ref{32}) grows more slowly than the previous ones. These differences depend once again on the structure of the coherent states taken into account. In addition, according to Eq.~(\ref{37.1}) the mean energy value depends as well on the magnetic field intensity as $\sqrt{B_0}$.%: the states in Eq.~(\ref{29.1}) have a closest structure to the standard coherent states, meanwhile the corresponding nonlinear ones in Eqs.~(\ref{28}) %, (\ref{30}) 
     %and (\ref{32}) are expressed in terms of monotonic functions $_pF_q$.
%   \section{Wigner function}
%   \begin{equation}\label{41.0}
%   W_\alpha(x,p)=\frac{1}{\pi}\int_{-\infty}^{\infty}\Psi^\dagger(x+y)\Psi(x-y)e^{2ipy}dy.
%   \end{equation}
%   Using the explicit form for $\psi_n(x)$, given in Eqn.~(\ref{8}), allows to find the Wigner function for each NLCS considered. In general, for two arbitrary exited states $\Psi_n$ of the Dirac-Weyl equation, we have
%   \begin{eqnarray}
%\nonumber   W_{n,m}(x,p)&=&\frac{1}{\pi}\int_{-\infty}^{\infty}\Psi^\dagger_n(x+y)\Psi_m(x-y)e^{2ipy}dy \\
%   &=&\frac{\omega^{1/2}}{2\pi}e^{-\xi^2-p^2}(-1)^m2^{m-1}(n-1)!(ip-\xi)^{n-m}\\
%\nonumber   &&\left[\frac{2n}{\sqrt{2^{n+m}n!m!}}L^{n-m}_{n}(2(\xi^2+p^2))-\frac{1}{\sqrt{2^{n+m-2}(n-1)!(m-1)!}}L^{n-m}_{n-1}(2(\xi^2+p^2))\right], 
%   \end{eqnarray}
%   where $L^m_n(x)$ denotes the associated Laguerre polynomials and $$\xi=\sqrt{\frac{\omega}{2}}\left(x+\frac{2k}{\omega}\right).$$
    \section{\label{sec:level4}Conclusions}   
		
Dirac electrons in graphene placed in homogeneous magnetic fields which are orthogonal to the material surface are ideal systems to start implementing the coherent states treatment in solid state physics. In particular, for constant magnetic fields the problem has been addressed for the first time quite recently \cite{wmg17}. In fact, in \cite{wmg17} the same physical configuration of this paper was considered, with the assumption that the magnetic field strenght is strong, in order that the Dirac electron stays always at the $n=0$ Landau energy level. On the other hand, in this article we are supposing that the magnetic field strenght is not strong, so that the state of the electron can be a coherent linear combination of all the eigenstates for the Landau energy levels. That is the reason why in this paper we required first to identify the appropriate annihilation and creation operators, in order to build then the coherent states as eigenstates of the former operator. Due to its non-uniqueness, however, it was possible to build different sets of coherent states. Although some of them could look similar to the standard coherent states for the harmonic oscillator, our graphene coherent states in general involve generalized hypergeometric functions. This dependence is more apparent when calculating the Heisenberg uncertainty relation for each set of this paper. This uncertainty achieves a minimum, equal to 1/4, for the coherent states of Eq.~(\ref{28}), since the ground state $\Psi_0$ is involved in this linear combination, while it reaches a maximum for the coherent states of Eqs.~(\ref{29.1}) and (\ref{32}), depending on the minimum excited state energy involved in the corresponding linear combination (see Figures \ref{fig:Graphene10}-\ref{fig:Graphene31}).

It is important to remark that, in a sense, the graphene coherent states remind the multiphoton coherent states \cite{perelomov72,barut71,buzek90,buzek901,sun92,jex93}, which appear from realizations of the Polynomial Heisenberg Algebras (PHA) for the harmonic oscillator \cite{fhn94,fnr95,fh99,carballo04,bermudez14,celeita16}. In that formalism, the Hilbert space decomposes as a direct sum of $m$ orthogonal subspaces, on each of which it is possible to construct the corresponding coherent states as superpositions of standard coherent states, while in the case of this paper the minimum energy states can be isolated from the remaining Hilbert subspace, depending on the values taken by $f(n)$.

On the other hand, the analysis of the probability density allows to characterize some physical properties of the graphene coherent states. This function indicates that the description for these states remains simple for finite $r$, whatever the value of the parameter $\theta$ is. However, the probability density reaches a maximum whose position along the $x$ axis actually depends on $\theta$ (see Figures \ref{Den1}-\ref{Den3}). Meanwhile, the behavior of the mean energy value suggests the possibility of using the graphene coherent states in semi-classical treatments.
  
Finally, it is important to stress that the non-uniqueness of the annihilation operator leaves open the possibility of exploring more complicated expressions for this operator. As a consequence, plenty of new sets of coherent states can be generated; some of them could be more useful than others for describing interesting physical phenomena in graphene and other carbon allotropes (see e.g. \cite{jknt13,jkn14,jt16}).

\end{document}